\documentclass[letterpaper,twocolumn,english,aps,pre,showpacs]{revtex4-1}
\usepackage[T1]{fontenc}
\usepackage[latin9]{inputenc}
\usepackage{amsmath}
\usepackage{amssymb}
\usepackage{graphicx}

\allowdisplaybreaks[1]

\makeatletter

\usepackage{times}

\makeatother

\usepackage{babel}
\begin{document}

\title{Discrete-attractor-like Tracking in Continuous Attractor Neural Networks}

\author{Chi Chung Alan Fung and Tomoki Fukai}

\affiliation{RIKEN Center for Brain Science, Hirosawa 2-1, Wako City, Saitama
351-0198, Japan}
\begin{abstract}
Continuous attractor neural networks generate a set of smoothly connected attractor states. In memory systems of the brain, these attractor states may represent continuous pieces of information such as spatial locations and head directions of animals. However, during the replay of previous experiences, hippocampal neurons show a discontinuous sequence in which discrete transitions of neural state are phase-locked with the slow-gamma (30-40 Hz) oscillation. Here, we explored the underlying mechanisms of the discontinuous sequence generation. We found that a continuous attractor neural network has several phases depending on the interactions between external input and local inhibitory feedback. The discrete-attractor-like behavior naturally emerges in one of these phases without any discreteness assumption. We propose that the dynamics of continuous attractor neural networks is the key to generate discontinuous state changes phase-locked to the brain rhythm. 
\end{abstract}


\maketitle
Ample evidence shows that the hippocampus performs sequence processing during the acquisition, consolidation and retrieval of memory. The latter characteristic has been extensively studied in spatial navigation tasks, in which hippocampal neurons in awake \citep{Kudrimoti1999} and sleep states \citep{Pavlides1989, Skaggs1996} replay the firing sequences that were exhibited during the preceding spatial experiences. These replay events are internally generated by the hippocampal circuits and believed to be part of neural processes for memory consolidation \citep{Carr2011, Eschenko2008, Girardeau2009}. On the other hand, the hippocampus has long been thought to operate as an attractor neural network in which episodes may be encoded into fixed-point attractors \citep{Knierim2016}, and this hypothesis also receives some support from experiment. However, how the seemingly different dynamical characteristics, attractor dynamics and sequence generation, emerge and cooperate in memory processing remain a mystery.

Recently, this question was addressed in the activity of CA1 neurons during a spatial memory task \citep{Pfeiffer2015}. Results of this experiment revealed that replay sequence during sleep, which represents sequence of the spatial locations visited previously by the animal, is discontinuous. Unlike smooth sequential firing during exploration in awake states, the sequence exhibited abrupt jumps between the decoded locations during replay events. Furthermore, these jumps were phase-locked to the slow-gamma oscillation of the local field potentials. In contrast, the hippocampal circuit showed attractor-like behavior during the replay of disconnected locations. These results suggested that slow decoding of accurate spatial information and fast movements between the remembered locations are alternated during spatial exploration. 

However, the underlying mechanisms of this oscillatory coding remain unknown. Here, we propose a continuous attractor neural network (CANN) to account for the emergence of discretized local attractors. The CANN is a family of neural field models that can support a set of continuously connected attractor states, which generally form bump-shaped functions in the space of preferred stimuli \citep{Amari1977, Wu2005}. The CANN models have been used to describe the tuning curves (i.e., place fields) of hippocampal place cells \citep{Samsonovich1997}, orientation tuning in visual cortex \citep{Ben-Yishai1995}, the direction of object movement in the middle temporal cortex \citep{Albright1984} and head direction in the entorhinal cortex \citep{Zhang1996}. In addition, the pairwise correlation predicted by CANN has been observed in experiment \citep{Wimmer2014}. By using a perturbative approach to the neural field dynamics of CANN \citep{Fung2008, Fung2010}, we demonstrate that the network model shows a phase transition from a continuous attractor state to discontinuous attractor states as the speed of an external stimulus is increased. This transition gives a discrete-attractor-like behavior similar to the experimental observation.

\begin{figure*}[t]
\includegraphics[width=1\linewidth]{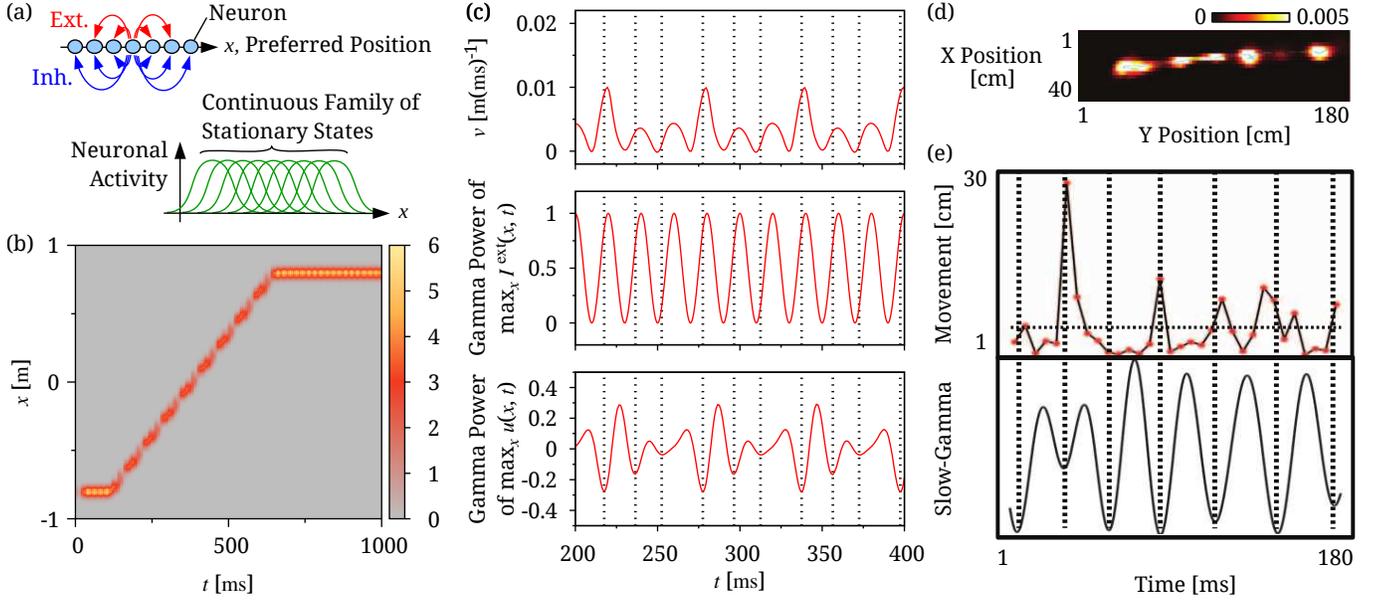}
\caption{\label{fig:Fig01} Basic response properties of CANNs. (a) In CANNs, neurons with similar preferred locations are mutually connected. Excitatory connections have a narrower range than inhibitory connections. This architecture allows the network to support a continuous family of stationary states. (b) Neural responses to a moving input are shown. Parameters: $\tilde{k}=1.0$, $a=0.02$ m, $\tau=2$ ms, $\tilde{v}= v/a=0.15$ (ms)\textsuperscript{-1} and $\tilde{A}_0 = 0.5$ (c) The instantaneous speed of the center of the localized activity profile (upper), magnitude of the external input function (middle) and the power of the gamma-band oscillation of neural field (lower) are shown (red curves) for the entire duration of moving input. Dotted lines indicate the troughs of gamma oscillations. Parameters: same as those in (b). (d) The location of a rat and (e) the slow-gamma power were decorded from hippocampal activity during replay. These panels were modified from \citep{Pfeiffer2015} with permission.}
\end{figure*}

We use the following version of a CANN model to describe the activity state $u\left(x,t\right)$ of neurons with preferred location $x$ at time $t$ \citep{Wu2005,Fung2008}:
\begin{align}
\tau\frac{du\left(x,t\right)}{dt}= & -u\left(x,t\right)+\rho\int dx^{\prime}J\left(x,x^{\prime}\right)u(x^{\prime},t)^{2}B(t)^{-1}\nonumber \\
 & +I^{{\rm ext}}\left(x,t\right),\label{eq:dudt_ori}
\end{align}
where $\rho$ is their density in the preferred location space and $\tau$ is the corresponding time constant. Excitatory coupling function between neurons at $x$ and $x^{\prime}$ is translational invariant and given as $J\left(x,x^{\prime}\right)=\tfrac{J_{0}}{\sqrt{2\pi}a}\exp[-\left|x-x^{\prime}\right|^{2}/(2a)^{2}]$, where $J_{0}$ controls the average magnitude of excitatory couplings and $a$ represents their average width. The function $B\left(t\right)=1+k\rho\int dx^{\prime}u\left(x^\prime,t\right)^{2}$ expresses a divisive inhibition \citep{Deneve1999,Wu2005} with a positive parameter $k$. The larger the value of $k$, the stronger the inhibition. External input to the network is given by a Gaussian function as $I^{{\rm ext}}\left(x,t\right)=A\left(t\right)\exp[-\left|x-z_{0}\left(t\right)\right|^{2}/(4a^{2})]$, when the current location of the animal is $z_{0}\left(t\right)$. Because external input influences network dynamics through the component projected onto the translational mode ($\partial u/\partial x$) of attractor states \citep{Fung2010}, the specific functional form of external input does not change the essential results. We assume that Eq. (\ref{eq:dudt_ori}) describes activity of hippocampal CA1 pyramidal cells. Local excitatory connections are less prominent in CA1 compared to CA3, but CA1 pyramidal cells are not devoid of recurrent synaptic connections \citep{Bezaire2013}. External input may arise from the entorhinal cortex or CA3 \citep{Yamamoto2017}. 

To simplify the parameter dependence of the model, we performed the following rescaling \citep{Fung2010}: 
\begin{alignat}{2}
u\left(x,t\right) & \rightarrow\tilde{u}\left(x,t\right) &  & \equiv\rho J_{0}u\left(x,t\right)\nonumber \\
k & \rightarrow\tilde{k} &  & \equiv8\sqrt{2\pi}ak\left(\rho\left.J_{0}\right.^{2}\right)^{-1}\label{eq:rescale}\\
A\left(t\right) & \rightarrow\tilde{A}\left(t\right) &  & \equiv\rho J_{0}A\left(t\right).\nonumber 
\end{alignat}
We first study solutions to Eq. (\ref{eq:dudt_ori}) for a vanishing external input ($\tilde{A}\left(t\right)=0$). For $\tilde{k}\ge1$, we can show that $\tilde{u}\left(x,t\right)=0$ is the only stable fixed point solution. For $\tilde{k}<1$, Eq. (\ref{eq:dudt_ori}) has a stable fixed point $\tilde{u}_{+}\left(x\right)$ and an unstable fixed point $\tilde{u}_{-}\left(x\right)$, where $\tilde{u}_{\pm}\left(x\right)=\sqrt{8}[(1\pm\sqrt{1-\tilde{k}})/\tilde{k}]\exp[-\left|x-c\right|^{2}/(4a^{2})]$ with $c$ being an arbitrary constant. The continuous family of stationary states is schematically illustrated in Fig. \ref{fig:Fig01}(a).
Now we turn to solutions for non-vanishing input. Owing to local excitatory connections and widely spread inhibition, network activity tends to form a localized profile, even though $\tilde{k}\ge 1.0$. A solution to Eq. (\ref{eq:dudt_ori}) is presented in Fig. \ref{fig:Fig01}(b). For comparison with experiment \citep{Pfeiffer2015}, we set as $a=0.02$ [m] and assumed an oscillatory input within the slow-gamma band $\tilde{A}\left(t\right)=\tilde{A}_{0}\times\left[\cos\left(2\pi ft\right)+1\right]$ with the amplitude and frequency of input set as $\tilde{A}_{0}=1$ and $f=50\ {\rm [Hz]}$, respectively. To mimic the observed replay sequence, we moved the input at a speed of $-0.8$ [m] to $+0.8$ [m] after an initial transient period of 100 ms. The evoked activity faithfully tracked the movement of external input, exhibiting fluctuations in the amplitude. 

To explore the biological relevance of the network behavior, we calculated the instantaneous speed of localized activity profile, $v(t) = dz(t)/dt$ with $z(t) =  \int x u\left(x,t\right)dx/\int u\left(x,t\right) dx$, in the upper panel of Fig. \ref{fig:Fig01}(c), which shows discrete-attractor-like behavior during tracking. The time course of the magnitude of the external input was shown in the middle panel for reference. We further examined the phase locking phenomenon between the gamma rhythm of the ``local field potential" and the change of the decoded location, where the amplitude of neural field $\tilde{u}\left(x,t\right)$ was used to address the ``local field potential" (LFP). Time evolution of the gamma-band power of $\tilde{u}\left(x,t\right)$ is presented in the lower panel of Fig. \ref{fig:Fig01}(c). From the two figures (upper and lower panels in Fig \ref{fig:Fig01}(c)), we can see that the peak speed of activity packet is phase locked to the troughs of gamma-band LFP oscillation in the model. We may compare these results with experimental observations. Figure \ref{fig:Fig01}(d) shows neuronal activity recorded from the rat hippocampus during replay \citep{Pfeiffer2015}. As in Fig. \ref{fig:Fig01}(b), the neuronal activity superimposed onto the decoded spatial locations was clearly discretized. Phase-locking similar to that in Fig. \ref{fig:Fig01}(c) also occurred between rat's movement and the slow-gamma LFP oscillation (Fig. \ref{fig:Fig01}(e)). 

At a first glance, the discretized response of the model looks trivial due to the presence of oscillating external drive. However, as demonstrated below, this phenomenon emerges from internal network dynamics but not from the oscillatory drive per se. To understand how the dynamics of the continuous attractor model generate discretized responses, we analyzed the model using the perturbative method developed in \citep{Fung2008, Fung2010}. In this approach, we construct a perturbative solution to the model by using a series of orthogonal functions.

\begin{figure}[t]
\includegraphics[width=1\linewidth]{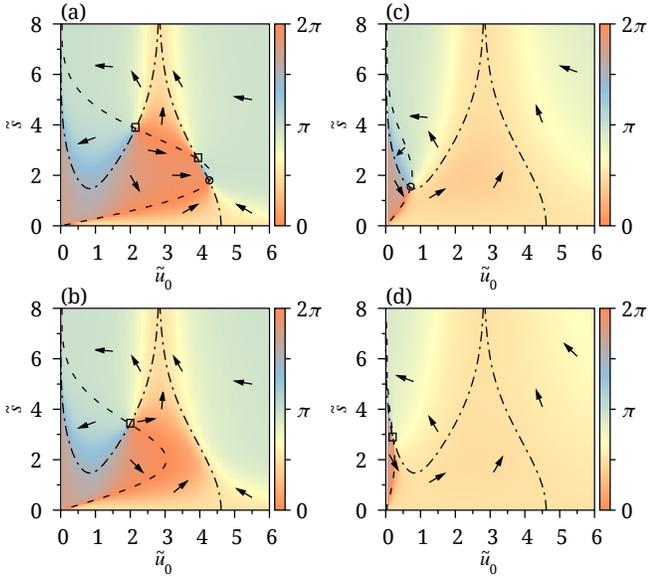}

\caption{\label{fig:plane_phase} Plane-phase diagrams for different parameter settings. Here, $\tilde{k}=1.0$ and $\tilde{A}=0.5$. (a) Plane-phase diagram shows three fixed-point solutions for $\tilde{v}_0=0.07$. Dashed and dot-dashed curves indicate $\tilde{s}$-nullcline and $\tilde{u}$-nullcline, respectively. (b) Plane-phase diagram gives an unstable fixed point for $\tilde{v}_0=0.1$. (c) Plane-phase diagram gives a stable fixed point for $\tilde{v}_0=0.4$. (d) Plane-phase diagram yields an unstable fixed-point for $\tilde{v}_0=1.2$. Arrows: visualized tendencies of nearby regions. Symbols: Circles: stable fixed point. Squares: unstable fixed point.}
\end{figure}

For mathematical simplicity, we assume that the localized activity profile can be approximated by a Gaussian function: $\tilde{u}\left(x,t\right)=\tilde{u}_{0}\left(t\right)\exp[-\left|x-z\left(t\right)\right|^{2}/(4a^{2})]$. Substituting this expression into Eq. (\ref{eq:dudt_ori}), 
rescaling the variables according to Eq. (\ref{eq:rescale}), and considering orthogonality of functions, we have
\begin{align}
\tau\frac{d\tilde{u}_{0}\left(t\right)}{dt} & = -\tilde{u}_{0}\left(t\right)+\frac{1}{\sqrt{2}}\frac{\tilde{u}_{0}\left(t\right)^{2}}{1+\frac{1}{8} \tilde{k}\tilde{u}_{0}\left(t\right)^{2}}+\tilde{A}\left(t\right)e^{-\frac{1}{8}\tilde{s}^{2}}, \label{eq:dtilde_u}\\
\tau\frac{d\tilde{s}\left(t\right)}{dt} & = \tau\tilde{v}_{0}-\frac{\tilde{A}\left(t\right)}{\tilde{u}_{0}\left(t\right)}\tilde{s}e^{-\frac{1}{8}\tilde{s}^{2}}, \label{eq:dtilde_s}
\end{align}
where $\tilde{s}=\left(z_{0}-z\right)/a$ and $\tilde{v}_{0}=\left(dz_{0}/dt\right)/a$ are the separation of activity and the (constant) speed of sensory input, respectively. 

Let $\left(\tilde{u}_{0}^{*},\tilde{s}^{*}\right)$ be a fixed point solution to Eqs. (\ref{eq:dtilde_u}) and (\ref{eq:dtilde_s}) for constant $\tilde{A}\left(t\right)$, namely, when the external input is a moving Gaussian packet without oscillatory modulations. Figure \ref{fig:plane_phase} shows the nullclines of Eqs. (\ref{eq:dtilde_u}) and (\ref{eq:dtilde_s}), in which a fixed-point solution corresponds to an intersection of the nullclines. The color code indicates the angle of $\left(d\tilde{u}_{0}(t)/dt\ /\ \tilde{u}_{0}(t),d\tilde{s}(t)/dt\ /\ \tilde{s}(t)\right)$ relative to the horizontal axis in a counterclockwise direction. Four possible scenarios exist depending on parameter values. In Fig. \ref{fig:plane_phase}(a), three fixed points exist, but only the solution with the smallest $\tilde{s}^{*}$ is stable. In Fig. \ref{fig:plane_phase}(b), the value of $\tilde{v}_{0}$ is increased such that only a fixed point with the largest $\tilde{s}^{*}$ may survive. This solution, however, is unstable. In Fig. \ref{fig:plane_phase}(c), the value of $\tilde{v}_{0}$ is further increased and that of $\tilde{u}_0^{*}$ is decreased, as expected from Eq. (\ref{eq:dtilde_s}), making $\left(\tilde{u}_{0}^{*}, \tilde{s}^{*}\right)$ a stable fixed point. In Fig. \ref{fig:plane_phase}(d), the fixed point eventually turns unstable and $\tilde{s}\left(t\right)$ diverges.

\begin{figure}[t]
\includegraphics[width=1\linewidth]{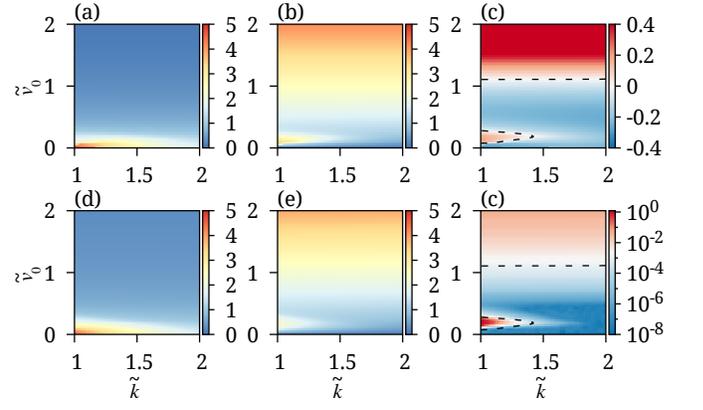}

\caption{\label{fig:measure_vs_theory} Parameter dependence of localized activity profiles. (a) $\tilde{u}_{0}^{*}$ of the fixed-point solutions to Eqs (\ref{eq:dtilde_u}) and (\ref{eq:dtilde_s}). (b) The value of $\tilde{s}^{*}$ in fixed-point solutions. In (a) and (b), only the fixed-point solutions with the smallest $\tilde{s}^{*}$ are presented if multiple fixed-point solutions exist. (c) Real part of the eigenvalues of the linearized system around the fixed-point solutions shown in (a) and (b). Dashed curves in (c) and (f) show contours on which the real part vanishes. (d) $\left\langle \max_{x}\tilde{u}\left(x,t\right)\right\rangle _{t}$ measured from simulations of Eq. (\ref{eq:dudt_ori}). (e) $\left\langle \tilde{s}\left(t\right)\right\rangle _{t}$ measured from the simulations. (f) Standard deviation of the measured $\tilde{s}\left(t\right)$. }
\end{figure}

We further studied the behavior of the network by solving Eqs. (\ref{eq:dtilde_u}) and (\ref{eq:dtilde_s}) in a broader range of parameter values. Figure \ref{fig:measure_vs_theory}(a) shows $\tilde{u}_0^*$ generally decreases with increases in $\tilde{v}_{0}$. If, however, $1<\tilde{k}\lesssim1.5$, as $\tilde{v}_{0}$ is increased from zero, the corresponding $\tilde{s}^*$ first increases but then decreases, taking a local maximum at some speed (Fig. \ref{fig:measure_vs_theory}(b)). This shows the presence of a parameter region in which the average separation is larger compared to the neighboring regions. A linear stability analysis was preformed around solutions to Eqs. (\ref{eq:dtilde_u}) and (\ref{eq:dtilde_s}) and the real part of the lowest eigenvalue is shown in Fig. \ref{fig:measure_vs_theory}(c) for various parameter values. The fixed-point solution is unstable in the parameter regions in which the real part is positive. 

In Fig \ref{fig:measure_vs_theory}(d)-(f), numerical simulations of Eq. (\ref{eq:dudt_ori}) were performed to confirm the above predictions of the reduced system defined by Eqs. (\ref{eq:dtilde_u}) and (\ref{eq:dtilde_s}). In (d), the peak height of localized activity profile $\left\langle \max_{x}\tilde{u}\left(x,t\right)\right\rangle _{t}$ averaged over $z_{0}\left(t\right)\in\left(-0.4\ {\rm m},0.4\ {\rm m}\right)$ was presented, while the temporal average of separation $\left\langle \tilde{s}\left(t\right)\right\rangle$  was shown in (e). They are comparable to the fixed-point solutions presented in panel (a) and (b), respectively. In stead of linear stability analysis, we evaluated the standard deviation of $\tilde{s}\left(t\right)$ in Fig. \ref{fig:measure_vs_theory}(f). By comparing this and Fig. \ref{fig:measure_vs_theory}(c), we find that the fixed point solution is unstable in the parameter regions where the standard deviation is larger than the minimum separation between neurons in the original model. The results indicate that the reduced system well replicates the dynamics of the original system. 

\begin{figure}[t]
\includegraphics[width=1\linewidth]{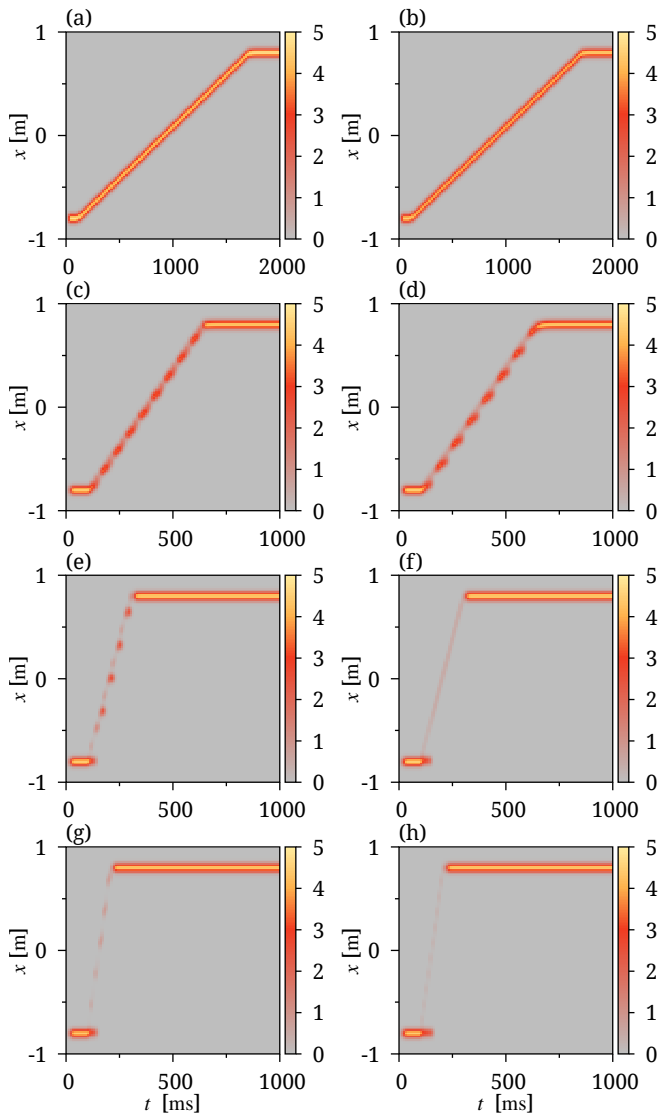}

\caption{\label{fig:more_sim} Similarity in network responses between constant and oscillating inputs. Panels in the left column show responses to various constant inputs with the amplitude $\tilde{A}=0.5$, while those in the right column show responses to various oscillating inputs with $\tilde{A}_{0}=0.5$. The speed of input $\tilde{v}_{0} =$ 0.1 (a, b), 0.3 (c, d), 0.8 (e, f) and 1.5 (g, h).  Values of the other parameters were the same as in Fig. \ref{fig:Fig01}.}
\end{figure}

The above results on the stability of activity profiles were derived for continuous attractor networks receiving constant inputs. Importantly, however, we found that the networks respond similarly to oscillatory inputs if the values of some parameters are adequately replaced. For example, in the phase shown in Fig. \ref{fig:plane_phase}(a), the profile of $\tilde{u}\left(x,t\right)$ steadily tracked a constant  input moving slowly (Fig. \ref{fig:more_sim}(a)). The network exhibited a similar response to an oscillating input with the same average amplitude moving at the same speed (Fig. \ref{fig:more_sim}(b)).  If the speed of constant input was increased to achieve the phase in Fig. \ref{fig:plane_phase}(b), the network  shows similar oscillating responses to both constant (Fig. \ref{fig:more_sim}(c)) and oscillating  (Fig.\ref{fig:more_sim}(d)) moving input. Because the oscillation appeared without oscillatory drive, the internal network dynamics likely drove $\tilde{u}\left(x,t\right)$ to oscillate. 

If the speed of input was further increased, $\tilde{u}\left(x,t\right)$ in Fig. \ref{fig:plane_phase}(e) and (g) followed the moving inputs with almost vanishing amplitudes, because the intrinsic dynamics were not sufficiently fast to respond to the rapidly moving inputs. With an oscillatory input, $\tilde{u}\left(x,t\right)$ still showed oscillatory responses, but with much reduced amplitudes (Fig. \ref{fig:plane_phase}(f) and (h)) presumably the intrinsic dynamics has only minor effect to the behavior.

The oscillatory response reported here is a result of the competition between attractor dynamics and a moving external input. We note that the external input needs not be oscillatory because the oscillatory responses are owing to the intrinsic network dynamics. It was previously shown that continuous attractors exist only for an adequate range of inhibition \citep{Fung2008, Fung2010}. If the inhibitory feedback is too strong, only small-sized activity packets can exist in the network. However, these packets are unable to reflect any property of attractor dynamics, such as the presence of fixed points, because recurrent inputs from surrounding neurons are much inferior to the external input. Conversely, if the inhibition is too weak, excitatory interactions between neurons become too strong to stably maintain the activity packets. Our results indicate that the oscillatory coding of spatial information with discrete-attractor-like states \citep{Pfeiffer2015} is possible only if the strength of inhibitory feedback falls into a marginal range. A testable prediction is that these states are impaired by the partial suppression of inhibition without crucially disturbing spatial memory.

In summary, this study presents the neural mechanisms that possibly underlie memory processing through oscillatory information coding. Since the slow-gamma oscillation is thought to play an active role in memory retrieval and consolidation during sleep or immobility \citep{Carr2012, Montgomery2007, Johnson2007}. How the intrinsic network behavior may contribute to memory consolidation is open for future studies.

\begin{acknowledgments}
This work was partly supported by Grants-in-Aid for Scientific Research (KAKENHI) from MEXT (17H06036) and CREST, JST (JPMJCR13W1) to T.F. 
\end{acknowledgments}

\end{document}